\documentclass[10pt,letterpaper]{article}

\usepackage{amsthm,amsmath}

\usepackage[utf8]{inputenc} 

\usepackage{nameref,hyperref}

\usepackage{graphicx}
\usepackage{xcolor}
\usepackage{multirow}
\usepackage{xspace}
\usepackage{bold-extra}

\usepackage{tikz}
\usetikzlibrary{shapes}
\usetikzlibrary{shapes.multipart}
\usetikzlibrary{fit}

\usepackage{amssymb}
\usepackage{mathrsfs}

\usepackage{geometry}
\usepackage{pdflscape}

\usepackage{booktabs}
\usepackage{makecell}

\usepackage{todonotes}

\newcommand{\ie}{\mbox{\textit{i.e.}}}
\newcommand{\eg}{\mbox{\textit{e.g.}}}
\newcommand{\gO}{\textsc{gitOmmix}}

\begin{document}
\vspace*{0.35in}

\begin{flushleft}
{\Large
\textbf\newline{Enhancing Clinical Data Warehouses with Provenance and Large File Management: The gitOmmix Approach for Clinical Omics Data}
}
\smallskip

Maxime Wack\textsuperscript{1,2,3,4}, 
Adrien Coulet\textsuperscript{1,2},
Anita Burgun\textsuperscript{1,2,5},
Bastien Rance\textsuperscript{1,2,3,$\star$}\\

\bigskip

\text{\textsuperscript{1}} Centre de Recherche des Cordeliers, Inserm, Université Paris Cité, Sorbonne Université, Paris, France
\\
\text{\textsuperscript{2}} Inria Paris, Paris, France
\\
\text{\textsuperscript{3}} Department of Biomedical Informatics, Hôpital Européen Georges Pompidou, AP-HP, Paris, France
\\
\text{\textsuperscript{4}} Centre Hospitalier National d'Ophtalmologie des Quinze-Vingts, IHU FOReSIGHT, 75012 Paris, France
\\
\text{\textsuperscript{5}} Imagine Institute, Inserm UMR 1163, Université Paris Cité, Paris, France 
\\

\bigskip
\textsuperscript{$\star$} corresponding author: \texttt{bastien.rance@aphp.fr}

\end{flushleft}

\bigskip
\bigskip

\paragraph{Abstract}

\paragraph{Background}
Clinical data warehouses (CDWs) are essential in the reuse of hospital data in observational studies or predictive modeling. 
However, state-of-the-art CDW systems present two drawbacks. First, they do not support the management of large data files, what is critical in medical genomics, radiology, digital pathology, and other domains where such files are generated.
Second, they do not provide provenance management or means to represent longitudinal relationships between patient events.  
Indeed, a disease diagnosis and its follow-up rely on multiple analyses.
In these cases no relationship between the data (\eg, a large file) and its associated analysis and decision can be documented.
\paragraph{Method}
We introduce \gO, an approach that overcomes these limitations, and illustrate its usefulness in the management of medical omics data. 
\gO~relies on 
\textit{(i)} a file versioning system: git, \textit{(ii)} an extension that handles large files: git-annex, \textit{(iii)} a provenance knowledge graph: PROV-O, and \textit{(iv)} an alignment between the git versioning information and the provenance knowledge graph.
\newpage
\paragraph{Results}
Capabilities inherited from git and git-annex enable retracing the history of a clinical interpretation back to the patient sample, through supporting data and analyses.
In addition, the provenance knowledge graph, aligned with the git versioning information, enables querying and browsing provenance relationships between these elements. 
\paragraph{Conclusion}
\gO~adds a provenance layer to CDWs, while scaling to large files and being agnostic of the CDW system. 
For these reasons, we think that it is a viable and generalizable solution for omics clinical studies.\\


\bigskip
\bigskip

\section*{Graphical abstract }
\begin{figure}[h!]
\includegraphics[width=\textwidth]{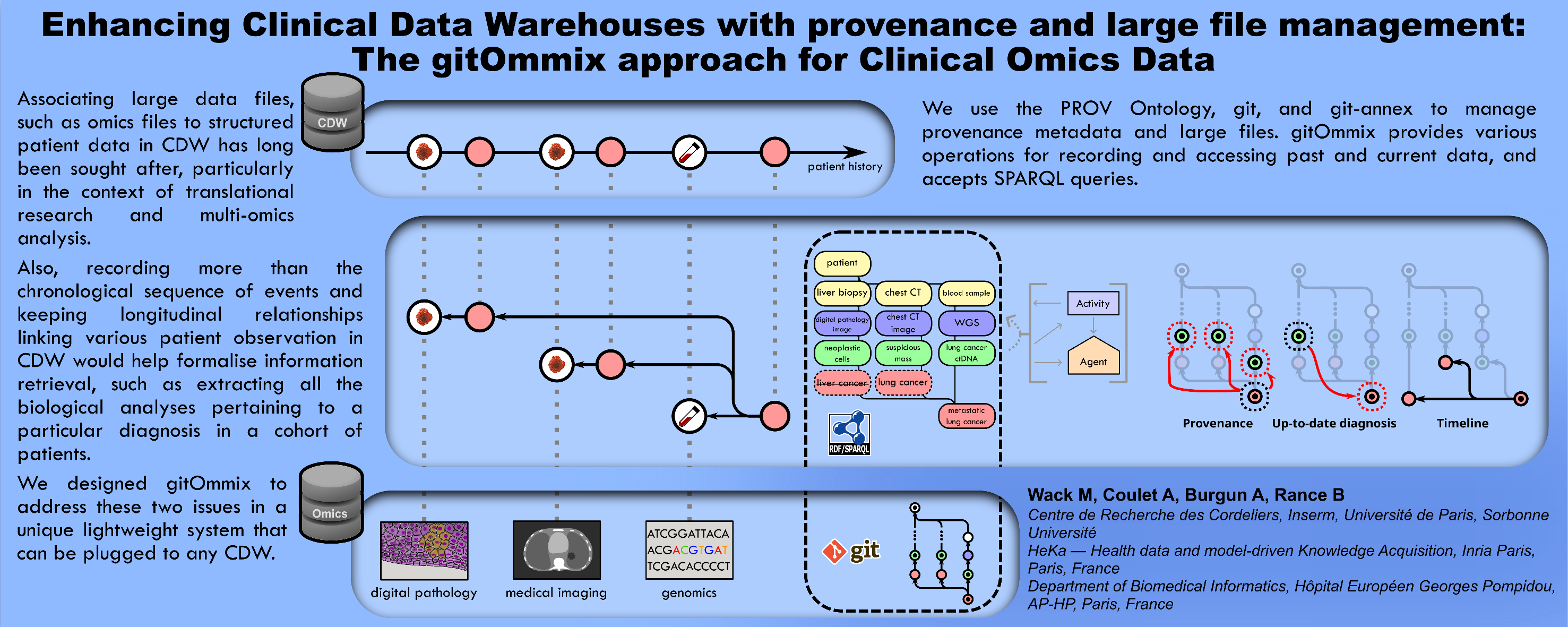}
\label{graphical_abstract}
\end{figure}


\newpage

\section{Introduction}

\paragraph{Background} 


With the rise of personalized medicine, patient omics data such as RNA or whole genome sequencing (WGS) enrich traditional clinical data, and by consequence find their place in electronic health records (EHR) and clinical data warehouses (CDW) \cite{canuel2015, trifan_patient_2019, robertson_it_2024}.
In this perspective, CDWs enriched with omics data offer an alternative to prospective cohorts for translational studies, \ie, studies typically searching for genotype -- phenotype associations such as genetic profiling of sub-groups of diseases or drug responses \cite{pendergrass2015, lyudovyk2019}.

CDW-based translational platforms present two main drawbacks that motivated this work.
The first drawback is the lack of management of data provenance. 
These platforms record patient events (such as observations, interventions, decisions) about patients in chronological order, but they seldom explicitly record historical relationships between these events.
Accordingly, the questions ``what are the observations that supported this decision?'', and inversely ``what decisions were made from this observation?'' can not be answered by these systems. The second drawback is the lack of management of large data files.
Relating a clinical decision, such as a diagnosis, to the content of a large file, such as the files of a whole-genome sequencing, remains difficult with CDWs.
However, these two functionalities are crucial for the management of clinical omics studies.

In computer science, \emph{data provenance} is defined as the documentation of where data comes from, and how it was transformed~\cite{bose2005}.
Among other aspects, provenance facilitates reproducibility in research ~\cite{goodman2016}, \ie~ the ability to obtain the same results by applying the same procedure to the same data ~\cite{benureau_re-run_2017}.
For this reason, standards and tools for data provenance have been developed~\cite{ikeda2013,lebo2013prov} and widely adopted in fields such as  bioinformatics~\cite{almeida2017,salazar2021}, but only parsimoniously diffused to medical informatics.
However, provenance and reproducibility are crucial for applications such as clinical decision support tools and their successful transfer to clinical practice.
This is particularly true when results are generated from prone-to-error biotechniques, potentially requiring several runs before confirming their validity.
One reason for the lack of data provenance management in medical informatics is its absence from CDW systems.
Most successful CDW models and their implementation, such as the i2b2 star model ~\cite{murphy_serving_2010}, the OMOP Common Data Model (CDM) ~\cite{voss2015omop}, or the eHOP model in France ~\cite{madec2019} are not supporting detailed provenance.




The \emph{management of large files}, \ie, larger than several hundred megabytes, or longer than 10 thousand lines, is also limited in CDWs.
This is mainly due to their use of relational database management systems, which are not designed to handle large files.
When such files are supported, they are usually stored aside from the CDW, and the CDW stores an unique reference such as an URL to the file. 
This is prone to inconsistencies and missing data, as file location and availability rely on independent file management systems, not synchronized with the CDW. 

\paragraph{Objective and motivation}

Our objective is to design an approach that allows storing large data files involved in clinical diagnoses and decisions, the relations between these data and diagnoses and decisions, as well as potential relations between clinical diagnoses and decisions; and to provide ways to query those relations and access the underlying data.


A common issue is to identify links between facts.
For example, identifying patients with a liver metastasis within a cohort of lung cancer patients.
This group of patients is not simply the set of patients with both diagnoses, as proof of the causal link between the original tumor and metastasis is necessary.
Our approach should enable querying that specific relation unambiguously as well as retrieving the supporting data (\eg, the digital pathology images of the primary lesion), and analyses (\eg, the search for variants in sequencing data associated with the disease progression tracking) for that relation.
Accordingly, it would help identifying patients satisfying complex inclusion criteria by querying the CDW in a more clinically meaningful way.

A core requirement of translational research is to access data obtained from high-throughput experiments and associated clinical data.
Our approach should enable finding all the observations related to a condition and its longitudinal follow-up, as well as retrieving the corresponding data.
More generally, it would allow the query of longitudinal information to access follow-up results or decision (or inversely to past data that motivated a decision).

\paragraph{Proposed solution} 

\gO~ allows provenance tracing, large file management, and the encoding of longitudinal relationships in CDWs, by combining:
\textit{(i)} the file versioning system git and its git-annex extension to manage large file histories,
\textit{(ii)} a knowledge graph to encode provenance metadata,
\textit{(iii)} a data model providing an alignment between these two systems, mapping data with metadata.

The rest of the article is organized as follow:
Section 2 presents the building bricks of our approach ; 
Section 3 presents \gO~itself ; 
Section 4 illustrates its use for the management of clinical omics studies. 

\section{Material}




\subsection{Semantic Web tools for data and provenance}

The Semantic Web proposes a set of standards and tools that facilitate sharing, linking, and processing data by associating them with formally defined semantics \cite{berners2001}.
This work relies on three Semantic Web standards: RDF (Resource Description Framework) \cite{W3C_RDF}, SPARQL (SPARQL Protocol And RDF Query Language), and PROV-O (PROV Ontology).
RDF, the Semantic Web standard for encoding knowledge graph, is a data model that represents data in the form \hbox{⟨subject, predicate, object⟩} triples, to describe a binary relation associating a subject and an object.
SPARQL is a query language for RDF knowledge graphs \cite{sparql_spec}.
%
PROV-O is a standard ontology recommended since 2013 by the W3C for the encoding of provenance metadata \cite{lebo2013prov}.
PROV-O is built around three main concepts: \texttt{Entities}, \texttt{Activities}, and \texttt{Agents}.
Entities represent physical or virtual objects, such as data sets or atomic elements of data.
Entities can be generated or modified by Activities.
Activities are realized by Agents.
Entities can also be directly attributed to Agents.
(Figure \ref{git2prov}a)

Adopting Semantic Web technologies provides additional tools contributing to the adherence to the FAIR principles \cite{wilkinson2016}.

\subsection{git and git-annex}

git is a distributed open-source file versioning system created in 2005 to support the Linux kernel development and now ubiquitously used in software development~\cite{git}.
git traces historical changes within files in a directory, called a \emph{repository}.
It uses a directed acyclic graph (DAG) structure, the \emph{git graph}, to record repository states, called \emph{commits}.
Because repositories are distributed and thus need to follow independent changes in various locations, branching and merging of histories is permitted, and is a core mechanism of collaborative development in software engineering. 
File additions, removals, or modifications are recorded in commits, which are accompanied by a \emph{commit message} describing the changes assigned to a \emph{commit author}.
Commits are uniquely identified in a repository by a cryptographic hash code, which can be seen as a signature of the content of the commit.
Any commit in the history of a repository and its associated files can be retrieved from the corresponding unique commit hash code.

git has originally been designed to trace changes in source code files, usually relatively small text files, but does not scale to large files.
git-annex, a third-party extension to git, has been created to overcome this limitation and handle large files \cite{gitannex}.
git-annex stores the designated files contents aside from the git repository and takes over the management of those files, while still recording the historical information by tracing a reference to the file within git.
It provides its own operations for adding and retrieving files, supporting a range of popular efficient file-hosting back-ends.
The stored reference is a cryptographic hash of the file content, making git-annex a \emph{content-addressable} file store: any change in a file has the consequence of modifying its cryptographic hash, enabling the unique identification of multiple versions of the same file.



\section{Methods - \gO}

We designed \gO~with three main components:
\begin{itemize}
    \item a \emph{data model} that records and semantically links clinical data and decisions that are related in term of provenance,
    \item a \emph{system} that traces changes in data, pointing at their up-to-date clinical interpretations,
    \item an \emph{association} between the data model and the system to ensure a progressive encoding of data provenance, at the time of data changes.
\end{itemize}
We defined a set of operations to build, manage, and query patient data history represented with the \gO~data model.

\emph{The \gO~data model} uses the PROV-O concepts of \texttt{Entities}, \texttt{Agents}, and \texttt{Activities}, and various possible relations between those concepts.

\begin{figure}
\centering
\includegraphics[width=8cm]{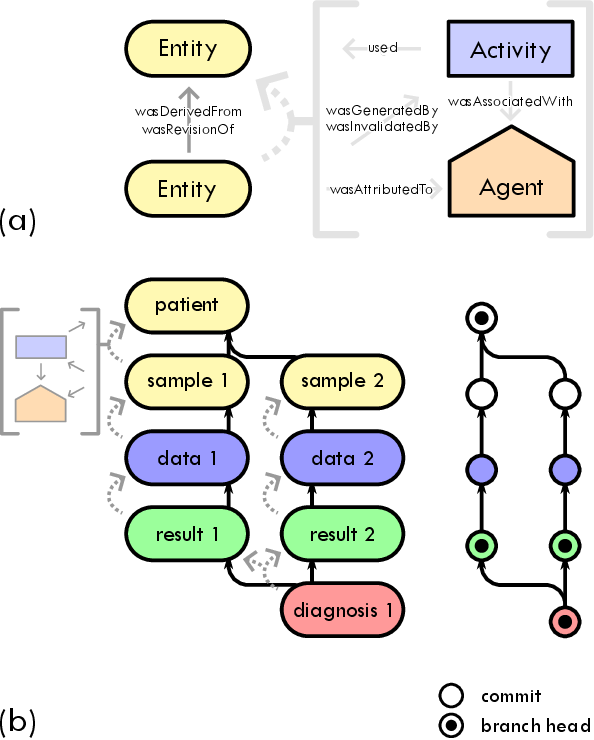}
\caption{The \gO~data model. (a) Derivation (or revision) relationship between two data elements of a CDW, represented with concepts and relationships of the PROV-O ontology. (b) Sequence of data elements deriving one from each other, as a repeat of the pattern showed in (a). The bottom left schema illustrates the correspondence between a sequence represented with PROV-O (on the left) and a git commit graph (on the right).  }\label{git2prov}
\end{figure}

We use them to represent clinical data and their provenance relations:
the \texttt{Agent} concept represents data providers, which can either be a human or a machine;
the \texttt{Activity} concept represents analyses, software runs or other methods that produce one data element;
the \texttt{Entity} concept represents any data element recorded in a CDW, associated or not with files. 
We extend entities into five subtypes: patients, samples, data, results, and diagnoses.
Patients and samples are considered as data elements because in the context of a CDW, they are indeed identifier of patients or samples.
Sample is a general naming encompassing identifiers of biological samples, but also of images or audio recordings.
Diagnoses can be any kind of clinical decision, but we restrict our study to diagnoses only, for simplicity. 

The most central relation of PROV-O, linking \texttt{Entities} together, is \texttt{was- DerivedFrom}. 
The relation \texttt{wasRevisionOf} 
is also used in the specific case of derivations that are data modifications. 
Figure \ref{git2prov}a illustrates these concepts and relations, and their use to represent data elements of a CDW and how one might derive from another. 
This derivation is a \textit{many-to-many} relation, as a sample can generate multiple data elements, and multiple results can lead to a single diagnosis. 
This relation between two entities is the atomic block that is repeated to build sequences of data elements derived from a patient, as illustrated in Figure \ref{git2prov}b. 
In our model, a diagnosis \texttt{wasDerivedFrom} a result, which \texttt{wasDerivedFrom} some data, which \texttt{wasDerivedFrom} a sample, which \texttt{wasDerivedFrom} a patient.
In this figure and throughout the rest of the article, we adopt the PROV-O prescribed shapes to distinguish between Entities, Activities and Agents.
In addition we use different colors to distinguish entities: blue for data, green for results, and red for diagnoses.
Other entities are kept blank.

For a more concrete example, a diagnosis of diabetes (an ICD10 code in the CDW), was derived from a laboratory result of high blood glucose concentration (a LOINC code), which in turn was derived from a blood glucose analysis (identified by an internal lab number), which was derived from a blood sample (a nursing procedure code).

In the specific case of modifications, updates or invalidations of diagnosis, results or data, the \texttt{wasDerivedFrom} relation is replaced by \texttt{wasRevisionOf}.
Providers and methods can optionally be added to further document the derivation relationship between entities. Accordingly, an entity $E_1$ \texttt{wasAttributedTo} a provider $P$, which \texttt{wasAssociatedWith} a method  $M$.
$E_1$ \texttt{wasGeneratedBy} $M$, which \texttt{used} a previous entity $E_2$.
For example, $E_1$ is a WGS assay attributed to a lab technician $P$, themselves associated with a short read sequencing $`$, which generated the sequence files $E_1$ using the patient sample $E_2$.

\begin{figure}
\centering
\includegraphics[width=14cm]{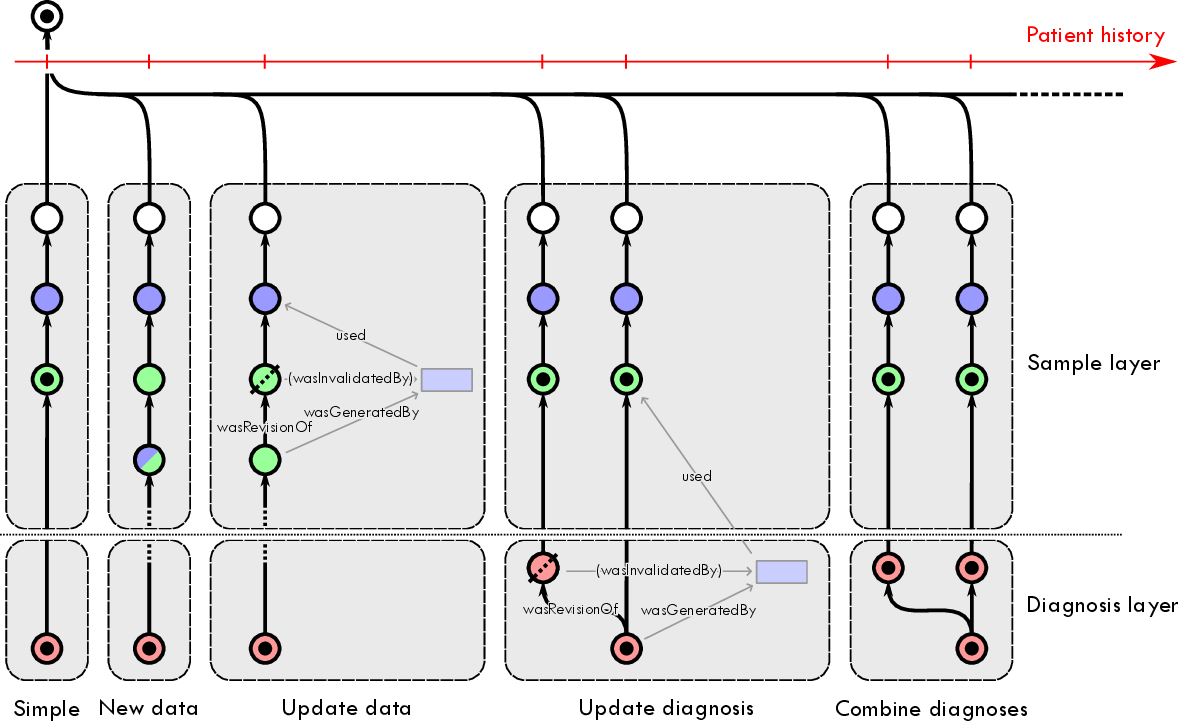}
\caption{Overview of the possible events composing a patient history in \gO. Nodes are commits, edges commit branching associated with a provenance type from PROV-O. Default type of relation is {\small \texttt{wasDerivedFrom}}, except when otherwise specified. Blue, green and red are data, results and diagnoses, respectively.}
\label{gitommix_overview}
\end{figure}

\paragraph{Implementing patient data history with git}

In \gO, we propose to rely on the git versioning system to trace patient data elements and associated clinical decisions.
Each patient is represented with its own single git repository, the git commit graph progressively built with new data, changes, and interpretations.
Small and large data files associated with patients observations are referenced and relationships between data elements and decisions are implemented using commits and git branching mechanisms. 
To facilitate maintaining this structure, we describe two ``layers": a \textit{sample} layer for the histories of data derived from every sample acquired from the patient (\eg, a biological sample, an image, or an audio file), and a \textit{diagnosis} layer for the relations between clinical decisions.
These two layers are illustrated in Figure \ref{gitommix_overview}.

The sample layer encompasses \emph{sample branches}.
Each new sample acquisition is materialized by a new git branch in the patient's git graph.
Each data and result derived from a sample is sequentially added to that sample branch as new commits.
Multiple revisions of these data or results can be added to a sample branch, possibly invalidating a previous version.\\
The second layer encompasses \emph{diagnosis branches}.
Clinical diagnoses are materialized by new branches on top of the sample branches, following a different construction rule.
Diagnosis commits can derive from one or multiple results (thus from multiple sample branches) using the git branch merging function.
Such a merge represents the joint contribution of multiple results to a single diagnosis.
Diagnoses can be further revised, combined, or invalidated by new merges of new results or diagnoses.

In git, branches are pointers to the latest commit in that branch, called the \emph{HEAD}.
In a sample branch, the HEAD points to the most up-to-date information and data related to that sample.
In a diagnosis branch, the HEAD points to the most up-to-date diagnosis.

\paragraph{Alignment between the data model and patient git graph}
Sequences of provenance relations represented with the data model can be aligned to git commit graphs, as illustrated in Figure \ref{git2prov}b. 
In this alignment, each PROV-O Entity corresponds to a commit in the git graph, and derivation relationships between Entities corresponds to parent-child commit relationships.

We implement this alignment by reusing the structure git offers to associate metadata to commits.
Indeed, every commit has an author, a date, and a message composed of a subject and a body.
We use the author and date to record the provider and date, respectively.
The message subject records the entity type and its id in the following form: \texttt{type:id} (\eg, \texttt{patient:123467890, diagnosis:ICD10\_I10}); and the message body records the associated metadata, encoded in turtle RDF.
For example, when adding a biopsy sample to a patient, the following RDF pattern is written into the commit message body:

\begin{verbatim}
:sample:{sampleId}
  a prov:Entity ;
  a :sample ;
  prov:generatedAtTime xsd:dateTime:{date} ;
  prov:wasDerivedFrom :patient:{patientId} ;
  rdf:label "{sampleId}" ;
  prov:wasAttributedTo :provider:{providerId} ;
  prov:wasGeneratedBy :method:biopsy .

:provider:{providerId}
  a prov:Agent ;
  a :provider .

:method:biopsy
  a prov:Activity ;
  a :method ;
  prov:startedAtTime xsd:dateTime:{date} ;
  prov:wasAssociatedWith :provider:{providerId} ;
  prov:used :patient:{patientId} .
\end{verbatim}

Using this commit metadata, the formal representation of provenance is preserved and closely associated with the corresponding data files, with the relationships between entities mirrored in the git graph structure.
Concatenating all the commit message bodies of a particular git history builds the corresponding RDF knowledge graph by incrementally adding nodes and relations.
The resulting knowledge graph has the advantage of offering query and reasoning facilities beyond those provided by git alone.  

\paragraph{Examples of patient data histories}

Figure \ref{gitommix_overview} illustrates the possible events in a patient history and their representation in \gO~.

The `Simple' box in Figure \ref{gitommix_overview} illustrates the trivial case of a diagnosis obtained from a single sample and a single biomedical analysis.
A sample branch is created (blank node); a data file is added with a commit (blue node); a result is added with a new commit (green node). 
Next, a diagnosis is added on top of the result by the creation of a new diagnosis branch (red node).
Target nodes are the HEAD of their respective branch.

The `New data' box in Figure \ref{gitommix_overview} illustrates the case of new data or results obtained from the same sample.
Those are added sequentially, accumulating the information produced from a single sample in the same history.
The graph structures of the RDF and git graphs can slightly differ here, as the git history stays linear while the RDF graph splits, as data always derives from the sample.

The `Update data' box in Figure \ref{gitommix_overview} illustrates the case of data or results updating or replacing previous ones Those are added sequentially, as in the previous case, with the use of the 
\texttt{wasRevisionOf}  relation intead of \texttt{wasDerivedFrom}.
In the case of invalidation, the invalidated entity is additionally documented with a temporal relationship \texttt{invalidatedAtTime}, and optionally with a relationship to the method it \texttt{wasInvalidatedBy}.
Note that multiple entities of the same kind can be invalidated at once by a single new entity.
In all those cases, accessing the repository ``at" a diagnosis gives access to all the files in the version that was used to lead to that diagnosis.

The `Update diagnosis' and `Combine diagnoses' boxes in Figure \ref{gitommix_overview} illustrate when a diagnosis emerges from the combination of multiple exam results, possibly originating from multiple samples, and from previous partial or erroneous diagnoses.
This is achieved in git using the merge operation to combine entities contributing to this diagnosis.
In both the update and combine cases a new diagnosis branch is created. 
The update diagnosis case is useful for aggregating more information related to a single diagnosis, or combining previous partial or symptomatic diagnoses into an etiological diagnosis.
The combined diagnoses case can be used to signal that multiple co-existing diagnoses are related through a syndromic diagnosis.
The new diagnosis can either stay unchanged when new analysis stay compatible with that diagnosis, or a different one derived from the additional information.

Figure S1 in Supplementary Data shows the actual git graph (a) and PROV graph (b) of the implementation of these operations in \gO~.
A script listing all the \gO~ commands needed to build this repository is provided with the software package.


\section{Results}

We implemented \gO~ as a set of operations and queries that can be called from a command line interface.
These commands are mostly abstractions over the underlying git commands managing the repositories.

\paragraph{Editing git graphs and associated provenance}
\gO~allows users to formulate simple 
commands such as ``add the NGS assay files of this patient's lung biopsy".
Each command triggers a series of git operations and enriches the commit metadata with provenance.  
Note that \gO~only enriches the git graph and metadata and does not suppress any of it, following the philosophy of version control systems. 
Accordingly, \gO~ provides operations to add, revise, or invalidate elements (patients, samples, data, results, diagnoses).
These operations can be combined to enable more complex ones, such as ``add the variant calling result derived from the NGS assay of this lung biopsy, and makes it revise the inconclusive pathology report produced earlier".

\paragraph{Queries with \gO}

\begin{figure}
\centering
\includegraphics[width=10cm]{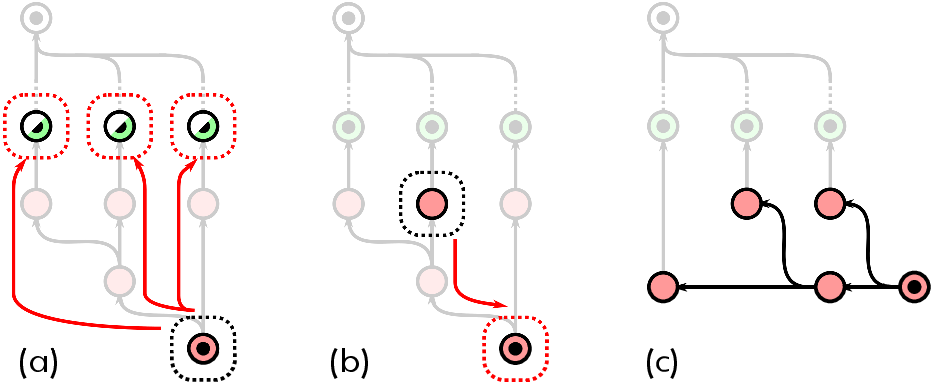}
\caption{The three types of queries enabled by \gO: (a) Provenance, (b) Up-to-date, and (c) Clinical timeline queries.}\label{gitommix_operations}
\end{figure}

\gO~ provides three main types of queries, illustrated in Figure \ref{gitommix_operations}.
(1) retrieving the provenance of any entity in the patient history.
This is supported by the simple fact that visiting a repository ``at" a commit accesses all the files accumulated up to that commit.
Large files stored in the annex are downloaded on demand only.
git logging facilities enable provenance to be further narrowed to specific entities, time periods, providers, etc.
Figure \ref{gitommix_operations}(a) illustrates this type of query listing all the data that contributed to an input diagnosis.

(2) retrieving the most up-to-date data, results, or diagnoses of a branch. 
As each sample and diagnosis is represented as a branch containing its whole history, navigating to its \emph{HEAD} returns the most recent version of data, results, and diagnoses, as illustrated Figure \ref{gitommix_operations}(b).
Chained with the first operation, this enables retrieving the additional data that documents the evolution of a diagnosis, \eg, a pathology report that documents the recurrence of a lung cancer.

(3) returning the timeline of a patient's successive diagnoses. Because the RDF representation of the provenance is represented in a piece-wise manner in the metadata of each entity, any subgraph of provenance can be built by concatenating a selection of pieces.
The resulting RDF can in turn be returned, queried in SPARQL, or used to produce graph visualizations.
The timeline operation displayed in Figure \ref{gitommix_operations}(c) is implemented with a SPARQL query that only selects diagnoses and diagnosis-diagnosis relations.

Indeed, in addition to these three common queries, \gO supports running arbitrary SPARQL queries on patients histories.

\begin{figure}
\centering
\includegraphics[width=\textwidth]{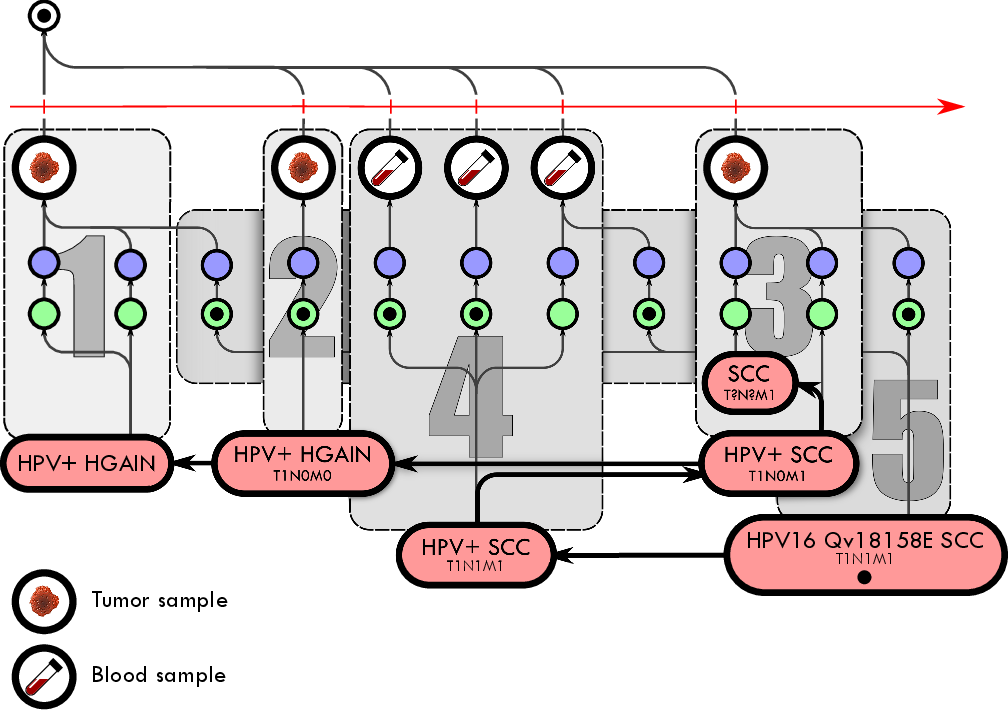}
\caption{Example of a patient's history in \gO:
(1) The patient is diagnosed with an HGAIN from the examination of a biopsy, with evidence of HPV6, 11, and 16 infection.
(2) The HGAIN is resected with free resection margins and as such is coded as T1N0M0.
(3) Later, a vertebral lesion is identified as a SCC metastasis of unspecific origin then linked to the HGAIN after evidence of HPV16 in the vertebral lesion.
(4) HPV16 ctDNA in increasing quantities is retrospectively detected on biobanked plasma samples.
(5) A specific variant of HPV16 in its episomal form is evidenced throughout all samples}
\label{gitommix_use_cases}
\end{figure}

\paragraph{A real-world clinical case}

We consider the previously published case report of a patient with a metastatic HPV-induced high grade anal intraepithelial neoplasia (HGAIN) \cite{pere_episomal_2021, veyer_hpv-circulating_2019}.
We schematized the git ommix graph associated with this patient history in Figure \ref{gitommix_use_cases}:
(1) The patient was diagnosed with an HGAIN during high-resolution anoscopy that led to the realization of a biopsy and pathologist confirmation of the diagnosis, completed with PCR identification of HPV6, 11, and 16.
(2) The HGAIN was subsequently surgically resected, with confirmed free resection margins.
(3) Two years later, the patient presented feverish back pains accompanied by weight loss, that were misdiagnosed at first.
After a couple of months and a new feverish episode, a new bone biopsy conducted the pathologist to conclude to a diagnosis of squamous cell carcinoma (SCC) metastasis.
Further investigations using immunohistochemical and PCR assays detected HPV16 DNA in the biopsy tissue.
This allowed linking the vertebral lesion to the HGAIN despite the absence of other signs of anal SCC contemporary to the metastasis diagnosis.
(4) Furthermore, as the patient had been participating in a research protocol that included the collection of plasma samples, those samples were retrospectively analyzed using quantitative digital droplet PCR, showing the presence of HPV16 circulating DNA in increasing blood concentration between the two diagnoses. 
(5) Further investigations using HPV capture and NGS showed that the exact same HPV16 subvariant was detected throughout all samples, from the initial HGAIN to the vertebral metastasis.

The actual PROV graph generated for this example is shown in Figure S2 of Supplementary Data.

\paragraph{Implementation}
An implementation of \gO~is available at {\small \url{https://www.github.com/gitOmmix/gitOmmix}} as Open Source Software.
It is implemented in bash and offers a user interface in the form of a command-line tool which includes help and auto-completion.
It relies on the command-line versions of git, git-annex, the rapper and roqet command-line tools from {\small \url{https://librdf.org}} for RDF file management, and graphviz \cite{graphviz} for graph visualization.

\section{Discussion}

\paragraph{Integration to clinical data warehouses}

\gO~is designed to enrich CDWs to enable the support of data provenance and large files.
Entities within a patient's repository are uniquely identified by an automatically generated identifier (a SHA1 hash of the git commit introducing the entity). 
Thus, the pair (patient id, unique hash) is an unambiguous reference to an object into an instance of \gO.
Accordingly, \gO~entities can be referenced in a CDW by associating the corresponding unique hashes to observations (\eg, the sample associated to the surgical procedure, the data file associated with a biological analysis, the full result report associated with an image exam, the diagnosis associated with a stay in the billing system).
In observation-based models such as i2b2, this can be achieved by re-using an existing column 
to store the associated hash.
In other scenarios where no free column is available, adding a specific column to the schema for this purpose is sufficient and does not interfere with the CDW. 
In CDWs using the OMOP CDM, although a similar mechanism could be used, a cleaner implementation using a new concept (\eg, ``gitommix\_hash") and a fact\_relationship linking the observation to its \gO~ hash would be preferable.
It is consequently possible to navigate back and forth between \gO~ and the CDW as necessary, filtering on patient identifier and commit hash in the observations table on the CDW side or targeting the commit hash on the \gO~ side.


\paragraph{Related works}
In a 2017 article \cite{murphy_combining_2017}, Murphy \textit{et al.} described three methods for combining clinical and genomic data within the i2b2 CDW.
The first one involves integrating genomic results as structured data using the Sequence Ontology.
The second one uses the i2b2/tranSMART platform with its \textit{ad hoc} ontology and data storage mechanism.
The third one uses a NoSQL database containing functionally annotated results, linked to i2b2 \textit{via} a custom i2b2 module.
All those methods involve transformations of the genetic results, making it impossible to access the primary data; and do not embed links between results, making it impossible to track longitudinal relationships between assays.
Our solution is architecturally similar to the third method, in that it adds an external layer to the CDW.
However, it is more independent as it does not rely on a specific CDW implementation and does not necessitate in-depth adaptation of CDW, but only relies on common tools.

\gO~ is compliant with any particular data schema or controlled vocabulary.
It could be added seamlessly to the first described method using the Sequence Ontology, linking each structured result to its source data.\\
Various initiatives have tentatively added structured representations of genomic data in the OMOP CDM, such as the genomic CDM (G-CDM) \cite{shin_genomic_2019}.
This complements \gO~ by enabling further structured representation of data hosted in \gO.

\paragraph{Limitations}

The current implementation of \gO~ is a proof of concept and for this reason presents some limitation.
First, it is local and centralized, and does not yet support the management of shared and remote repositories, as permitted by git and git-annex and intended for \mbox{\gO}.
Second, its interface is still rudimentary and limited to experienced users.


\paragraph{Advantages}

However, our solution has multiple advantages over previously described systems, by giving the ability to integrate arbitrary large data and by enabling the representation of relations between these data points.
It does so without relying on an entirely new paradigm around data representation in CDW, or needing heavy adaptations to the system currently in use.
It acts as a plug-in solution, agnostic from the CDW system in use, and is supported by standards and tools that were originally designed to address the specific issues at hand: keeping record of file histories, managing large files, and tracing provenance in a formal way.

\gO, as well as the tools it relies on, is open source backed by open standards.
This allows to readily benefit from the capabilities of these tools, enabling powerful file management, and clearly defined semantics, reasoning capabilities, and interoperability.
Although our system prescribes a general structure to its data model, it does not restrict the usage of additional features from the underlying systems.
For example, the base PROV-O triples generated for each entity can be supplemented at will to construct richer provenance graphs.
git-annex supports a diverse collection of file storage back-ends to host arbitrary large files, locally or remotely, on-premise or in the cloud, offering to benefit from efficient storage solutions.

Regarding scalability, as git-annex separates the management of large file from the management of the repository, repositories themselves stay very light in term of memory and thus responsive to query.
And because each patient exists as its own git repository, operations can inherently parallelized by running as many \gO~processes as needed.


\paragraph{Perspectives}

By enriching data with provenance metadata and enabling access to versioned source data files, \gO~ contributes to a better adherence to FAIR principles in the management of complex clinical data.
In particular, it ensures findability by assigning persistent and unambiguous identifiers, providing rich metadata, and a standard way to search within this metadata; accessibility by making source data available through git; interoperability by relying on standard knowledge representations; and reusability by adding detailed provenance and allowing access to all versions of the data.

For these reasons \gO~ allows reproducibility and consistency in  conducting translational studies, particularly retrospective studies based on large data that are more and more routinely collected during care.
In addition, it may also benefit clinical care as it documents clinical decisions explicitly and in a FAIR format \cite{robertson_it_2024}.

On the technical side, using established standards and tools allows for the addition of features supported by those tools.
For example, git enables authors to cryptographically sign their commits, which could be used to add a layer of security to the tracing of provenance.
git repositories can contain references to other git repositories using \textit{submodules}.
Using submodules could allow even richer provenance tracing by directly referencing the actual analysis code, pipeline, or tool at the version in which it was used to produce observations.




\section{Conclusion}

We introduce \gO, a relatively simple and lightweight system combining semantic web, file versioning, and content-addressable distributed file storage to represent and manage provenance and large source data in clinical data warehouses.
It includes all the functions required to build a patient's history graph and store associated files, navigate and query history using SPARQL, and retrieve the specific files related to any event.
We base our proposition on widely accepted systems and a model leveraging the shared DAG structure underlying these systems.

We provide a proof of concept implementation demonstrating feasibility and practical use of \gO, and illustrate its use with real-word use case about diagnosis based on clinical omics data. 
It is open to contributions and will be extended to support additional functions.

\section*{Abbreviations}

\noindent
CDW: Clinical Data Warehouse\\
ctDNA: circulating DNA\\
DAG: Directed Acyclic Graph\\
ddPCR: digital droplet Polymerase Chain Reaction\\
EHR: Electronic Health Record\\
FAIR: Findable, Accessible, Interoperable, Reusable\\
G-CDM: Genomic Common Data Model\\
HGAIN: High Grade Anal Intraepithelial Neoplasm\\
HPV: Human Papilloma Virus\\
i2b2: Informatics for Integrating Biology and the Bedside\\
ICD-10: International Classification of Diseases, 10th revision\\
LOINC: Logical Observation Identifiers Names and Codes\\
NLP: Natural Language Processing\\
NoSQL: non-SQL\\
OMOP CDM: Observational Medical Outcomes Partnership Common Data Model \\
OWL: Web Ontology Language\\
POC: Proof of Concept\\
PROV-O: Provenance Ontology\\
RDF: Resource Description Framework\\
RDFS: RDF Schema\\
SCC: Squamous Cell Carcinoma\\
SHA-1: Secure Hash Algorithm 1\\
SPARQL: SPARQL Protocol and RDF Query Language\\
WGS: Whole Genome Sequencing

\section*{Authors contributions}

\noindent MW: conceptualization, software, visualization, writing (original draft, review and editing).\\
AC: supervision, conceptualization, writing (review and editing).\\
AB: writing (review and editing), validation, funding acquisition.\\
BR: supervision, conceptualization, funding acquisition, writing (editing and review)

\section*{Funding resources}

We benefit from a government grant managed by the Agence Nationale de la Recherche under the France 2030 program, reference ANR-22-PESN-0007 ShareFAIR.

\section*{Acknowledgments}

Dr. Hélène Péré and Dr. David Veyer for the fruitful discussions about their inspiring clinical projects.
Linus Torvalds for creating linux and git, on which this contribution is based.

\bibliography{article}

\end{document}